\documentclass[a4paper]{article}
\usepackage[T1]{fontenc}
\usepackage{color}
\usepackage{graphics}

\makeatletter

\providecommand{\LyX}{L\kern-.1667em\lower.25em\hbox{Y}\kern-.125emX\@}

\newcommand{\lyxrightaddress}[1]{
  \par {\raggedleft \begin{tabular}{l}\ignorespaces
  #1
  \end{tabular}
  \vspace{1.4em}
  \par}
}

\usepackage[T1]{fontenc}

\makeatletter

\usepackage[T1]{fontenc}

\makeatletter

\usepackage[T1]{fontenc}

\makeatletter

\usepackage[T1]{fontenc}

\makeatletter

\usepackage[T1]{fontenc}


\makeatother
\makeatother
\makeatother

\makeatother

\begin{document}

\title{{\Large Baryon Transport in Dual Models and }\\
 {\Large the Possibility of a Backward Peak in Diffraction}\Large }

\author{{\normalsize Fritz W. Bopp}\thanks{
\noindent The speaker thanks the Organizing Committee for the interesting and
pleasant conference.
} {\normalsize }\\
 {\normalsize Universität Siegen, Fachbereich Physik,  D--57068 Siegen, Germany
}\\
{\normalsize presented at the conference }\\
{\normalsize CRIMEAN SUMMER SCHOOL-SEMINAR, Yalta , May 27 - June 4, 2000}\normalsize }

\maketitle
\vspace*{-7cm}

\lyxrightaddress{SI-00-7}

\vspace*{+5.5cm}

\begin{abstract}
We begin to briefly survey the experimental and conceptual side of baryon transfers
in particle scattering. A discussion of baryon transfers in heavy ion scattering
follows. It shortly reviews existing string model concepts, which were found
to be consistent with the data. With this motivation we turn to a more careful
consideration of the relevant topological structures. The baryon transfer is
associated with one of two possible cuts in baryonium exchange. From the color
structure the baryonium can be identified with an Odderon exchange. We conjecture
that the two Odderon cuts occur with an opposite sign and partially cancel.
As the Odderon is predicted to have a rather high trajectory it has to involve
small coupling constants. As this suppression is not anticipated for diffractive
processes a tiny observable backward peak is argued to occur in the initial
baryon distribution in massive diffractive systems. 
\end{abstract}

\section{Baryon transfer in particle scattering}

\textbf{Available experimental data}

To observe the dynamics of baryon transfers over large rapidity gaps the best
data involve the stopping of leading baryons to central rapidities. This requires
to somehow identify the initial baryons distribution. One has to rely on the
hypothesis - which presumably holds to a satisfactory degree - that the produced
``sea'' baryons and anti-baryons have an identical contribution and that a
simple subtraction yields the distribution of the net initial baryon charge. 

Unfortunately this excludes \( p{}\bar{p} \) scattering, where the sea-baryon
contribution cannot be determined from data. Unfortunately this precludes the
use of post ISR data excepting HERA and diffractive systems with sufficiently
large Pomeron-proton subenergies. The data are therefore typically quite old.
Available are spectra from meson-baryon processes (compiled in \cite{1}) and
from ISR (compiled in \cite{2}). At ISR a suitable combination of proton-proton
and proton-antiproton processes can be used to separate forward and backward
component of the incoming charge with a reasonable factorization assumption.
An old plot \cite{3} of the leading baryon charge shown below is found to be
consistent with a slope (in y) of \( \alpha _{Transfer}-\alpha _{Pomeron}=-1 \)
with large error. That the central data points are somewhat on the high side
can be taken as a hint of an eventual turnover to a flatter value. 

\vspace{0.3cm}
{\par\centering \resizebox*{0.9\textwidth}{!}{\includegraphics{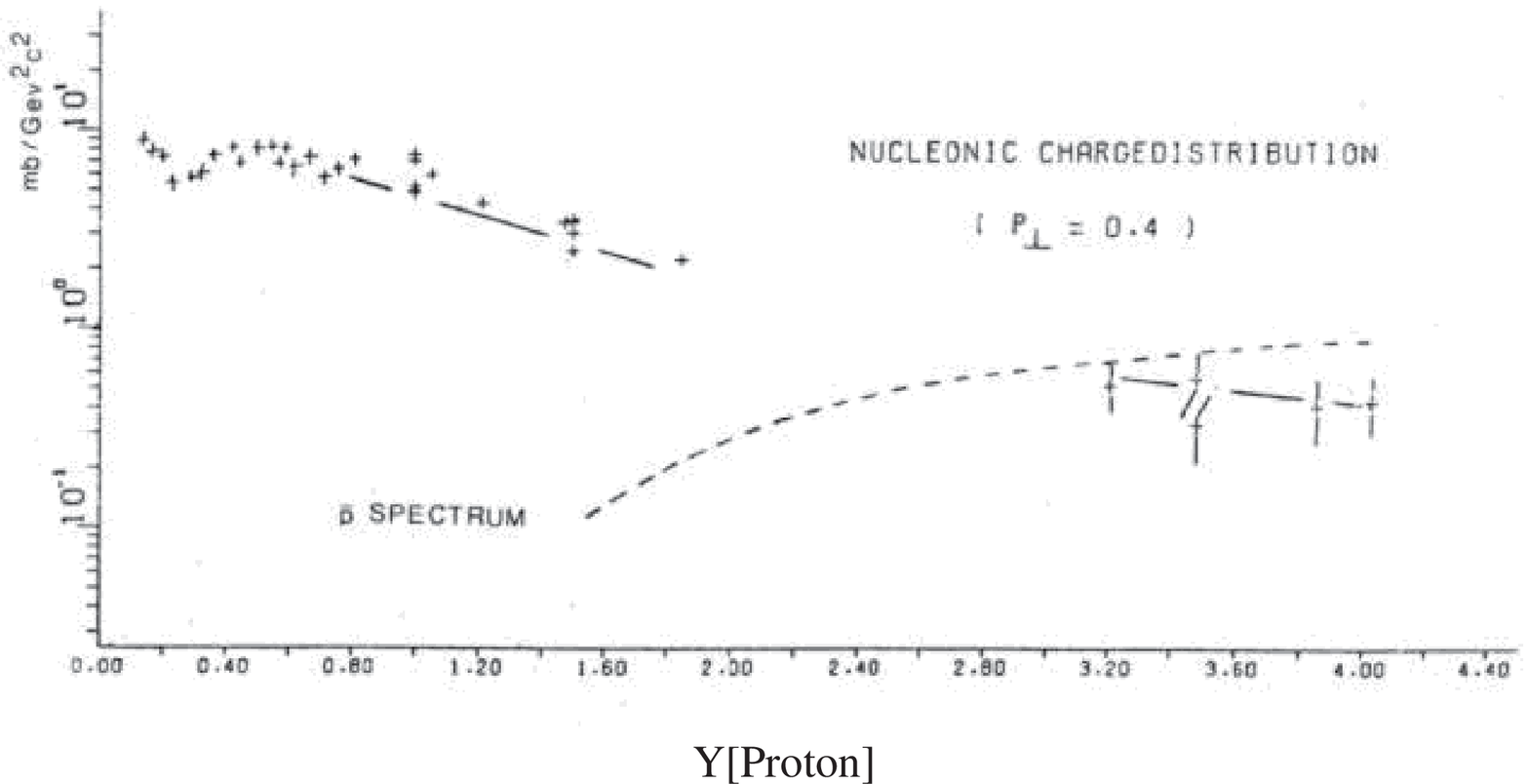}} \par}
\vspace{0.3cm}

To reduce the error one can integrate over the transverse momenta and to minimize
the systematical error one considers the ratio:

\[
A_{ISR}=\frac{\rho _{initial\, baryon\, charge}(y)}{\rho _{sea\, baryon\, charge}(y)}=0.39\pm 0.05,\: 0.33\pm 0.05,\: 0.23\pm 0.05\]
 given in a graph of \cite{4,5} for \( y=-0.4 \), \( y=0 \) and \( y=+0.4 \).
The central derivative of \( A_{ISR} \) obtains no contribution from the (symmetric)
sea-baryon distribution. Assuming the usual exponential distribution \( A\propto \exp [(\alpha _{Transfer}-\alpha _{Pomeron})y] \)
the quantity 
\begin{equation}
\frac{d/dy\, A_{ISR}}{A_{ISR}}|_{y=0}=\alpha _{Transfer}-\alpha _{Pomeron}=-0.49_{-0.37}^{+0.42}
\end{equation}
 just yields the slope. While not yet in contradiction with the initial intercept
the preferred value is now considerably less.

New preliminary data come from the H1 experiment at HERA \cite{6}. They observed
the initial baryons asymmetry at laboratory rapidities

\begin{equation}
A_{H1}=\frac{\rho _{initial\, baryon\, charge}(y)}{\rho _{sea\, baryon\, charge}(y)}=0.08\pm 0.01\pm 0.025
\end{equation}
 A simple extrapolation of the ISR value accounting for the larger rapidity
range spanned would have \char`\"{}predicted\char`\"{}: 

\begin{equation}
A_{H1}=0.061_{-0.046}^{+0.243}
\end{equation}
 Hence the H1 values lie within the expected range. The trajectories required
by the HERA ratio compared with its ISR value is now (see also \cite{7}) 
\[
\alpha _{Transfer}-\alpha _{Pomeron}=-0.4\pm 0.2\: .\]
 If the preliminary data are finalized, this confirms the flattening of the
trajectory.

\paragraph{Dual Topological picture }

The slowing down of the baryons is determined by the intercepts of the relevant
exchanges. The basic philosophy of the Dual Topological model \cite{8} in the
classification of such exchanges \cite{9} involves ``materializing'' or ``suppressed''
strings. ``Materializing'' means that the initial color fields are neutralized
by a chain of hadronizing \( qq \) pairs, ``suppressed'' means hadron-less
neutralization by an exchange of a single quark. It is well-known for the Pomeron
and the Reggeon exchanges where the cut Pomeron contains two materializing strings
while the cut Reggeon contains one suppressed and one materializing string.
As a general rule contributions with various suppressed strings have to be considered
as independent and additive. For each suppressed string an extra factor \( (\sqrt{1/M_{string}}) \)
appears and restricts the suppressed contribution to low energies or narrow
rapidity ranges.

For a nuclear exchange one starts with a completely suppressed exchange, i.
e. with the square of the quasi-elastic nucleon exchange amplitude
\begin{equation}
\alpha _{junction}^{III}-1=2(<\alpha _{Nucleon}>-1)=-2
\end{equation}
 known from elastic backward scattering. Each of the three exchanged valence
quarks can now be replaced by a ``materializing'' string. Corresponding to
three, two, one or zero strings there are four contributions with trajectories
spaced by one half. At considered energies the leading two of these ``baryonium''
trajectories with two and three hadronizing strings
\begin{equation}
\alpha _{junction}^{0}-1=-0.5,\: \alpha _{junction}^{I}-1=-1.0
\end{equation}
 will be relevant. They could be responsible for the initially steep (\( \approx -1.0 \))
and then possible flattening (\( \approx -0.5 \)) slope observed in the data
discussed above.

However the value of the final trajectory is rather uncertain. Values of \( \alpha _{junction}^{I}-1=-0.8\: ...\: 0 \)
were proposed in the literature \cite{10,11,12}. The correspondence to the
Odderon discussed below will give support to a flat value.

\paragraph{Implementation in Dual Parton model based Monte Carlo codes}

At present energies nonleading Regge exchanges will have no strong effect on
cross sections. To understand the final state structure local baryonium exchanges
within a global Pomeron exchange have to be considered. A factorization among
strings allows to ignore the quark string which is common to both trajectories
and the inclusion of such exchanges is therefore straightforward:

{\par\centering {\small \resizebox*{!}{5cm}{\includegraphics{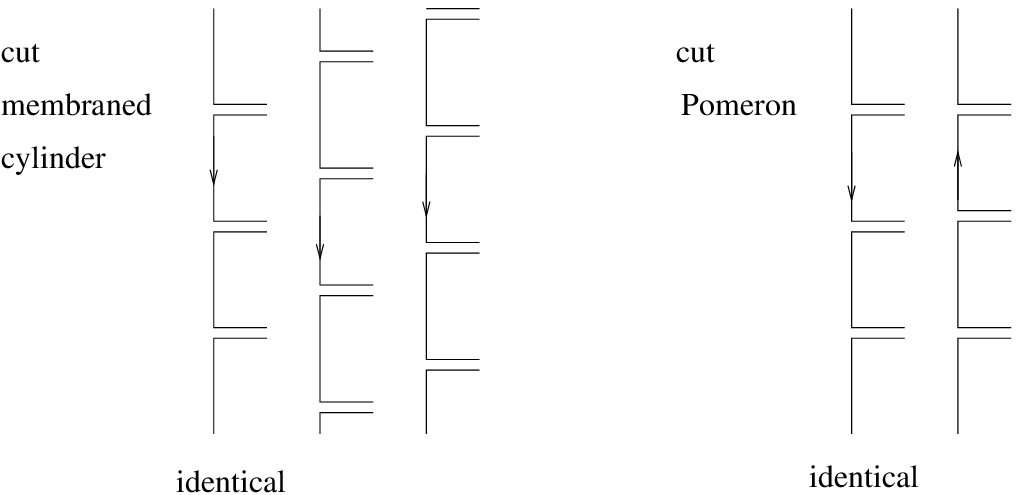}} }\small \par}

\noindent The transition of the remaining diquark string (baryonium remnant)
into an antiquark string (Pomeron remnant) can be implemented in a usual fragmentation
scheme by a suitable choice of the splitting function involving quark and diquark
transitions. It was implemented in most string models e.g. \cite{13,14,15}
and it is part of the JETSET program (as diquarks or as pop-corn mechanism \cite{16}).
Without relying on string factorization leading- and sea-baryon exchanges are
also implemented in HIJING/\( B\bar{B} \) \cite{17}.

\section{Baryon enhancement \\
 in dense heavy ion scattering}

\paragraph{Concepts for slowing-down initial baryons }

There are a number of conventional mechanisms enhancing baryon transfer and
central baryon production in multiple scattering processes for string models.
They are helpful in some regions but not enough to explain the large stopping
in heavy ion scattering \cite{18,19}.

To understand the data it seems necessary to include interplay of string if
they get sufficiently dense in transverse space. It was proposed that there
are new special strings \cite{20,21}. In contrast, we shall maintain here the
general factorization hypothesis between initial scattering in the quark phase
and the final hadronization within standard strings. 

Consider an incoming baryon. The usual Pomeron exchange in the Dual Parton model
leaves a quark and a diquark for the string ends. Diquarks are no special entities
and multiple scattering processes have no reason not to split them in a conventional
two Pomeron interaction \cite{22,7,23,24,25}. It is natural to expect that
diquark break-ups considerably slow down the baryons evolving. The probability
for such an essentially unabsorbed \cite{26} process is \cite{27,28,29,22}:
\begin{equation}
{[break\: up{]}/{[}no\: break\: up{]}\propto {[}cut\: Pomeron\: number{]}-1}
\end{equation}
 As required by the experimentally observed slowdown this is a drastic effect
for heavy ion scattering \cite{22} while for hadron-hadron scattering multiple
scattering are sufficiently rare to preserve the known hadron-hadron phenomenology.
How such processes are affected was considered numerically in \cite{25} and
no manifestly disturbing effects were found.

We emphasize that the behavior of the baryon quantum number slowed down by such
a break-up is not trivial. In topological models the baryon contains Y-shaped
color electric fluxes. Two Pomerons intercepting two different branches will
leave two ``free'' valence quarks and a valence quark connected with the vortex
line with the velocity of initial baryon which will subsequently form the end
of the strings. The energy distribution of quarks with vortex lines (or of the
fully separated vortex lines) in the structure function is a priory not known.

\paragraph{Special baryon transfers in the Topological model}

For a more detailed description of the slowing down we turn to the Dual Topological
model \cite{9} introduced above. A discussion of baryon transfers in such a
framework was recently given by Kharzeev \cite{30}. We will here emphasize
topological aspects.

In topological models a Pomeron exchange corresponds to a cylinder connecting
the two scattering hadrons. If one considers an arbitrary plane intersecting
this exchange the intersection of the cylinder is topologically a circle. More
specifically amplitudes with clockwise respectively anticlockwise orientation
have to be added or subtracted depending on the charge parity. The cylinders
or the circles therefore come with two orientations. This distinction is usually
not very important as it is always topologically possible to attach hadrons
in a matching way; except for \( C \)-parity conservation no special restrictions
result. 

Pomerons have a transverse extent and if they get close in transverse space
they should interact. Hadronic interaction is sufficiently strong to be largely
determined by geometry. It is therefore reasonable to expect that the coupling
does not strongly depend on the orientation as long as there is no mechanism
of suppression. 

The two distinct configurations lead to different interactions. Two Pomerons
with the same orientation can if they touch (starting locally at one point in
the exchange-channel time) shorten their circumference and form a single circle:

{\par\centering \resizebox*{0.5\textwidth}{!}{\includegraphics{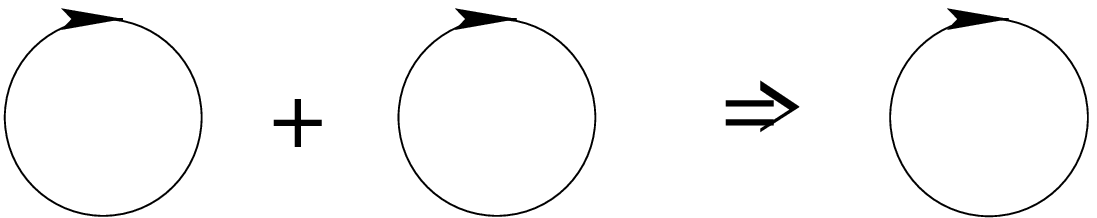}} \par}

\noindent This then corresponds to the usual triple Pomeron coupling experimentally
well known from diffractive processes. 

For two Pomerons with opposite orientation the situation is more complicated.
Like for soap bubbles the two surfaces which get in contact can merge and form
a single membrane. The joining inverts the orientation of the membrane. On the
intersecting plane one now obtains -- instead of the single circle -- three
lines originating in a vortex point and ending in an anti-vortex point as shown
below: 

{\par\centering \resizebox*{0.5\textwidth}{!}{\includegraphics{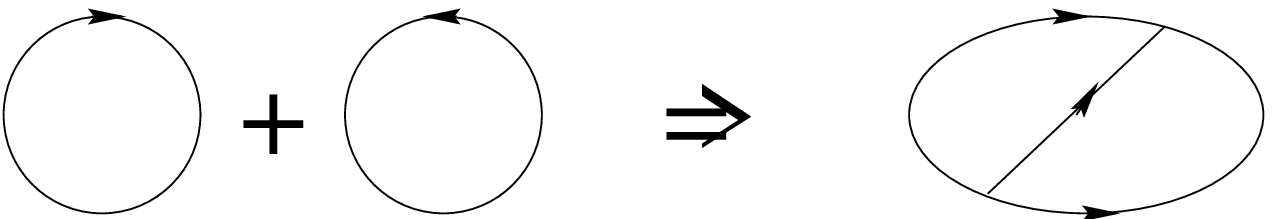}} \par}

\noindent Lacking a topological name for the object the term membraned cylinder
will be used in the following.

How do this membraned cylinder contribute to particle production? Similar to
the triple Pomeron case there are three different ways to cut through a membraned
cylinder:

{\par\noindent \centering \resizebox*{0.22\textwidth}{!}{\includegraphics{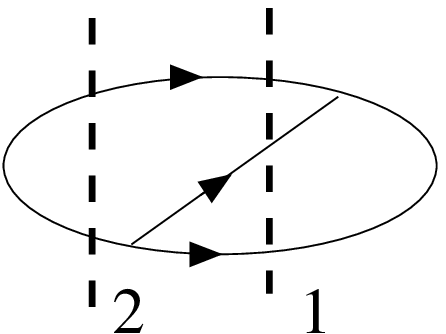}} \par}

\noindent The cut numbered 1 which also intersects the membrane has vortex lines
on both sides. They present a topological description of the baryon transfers
considered above. By symmetry they contribute with a positive sign. Cuts which
intersect only two sheets numbered 2 contribute to the two string contribution.
Their sign is unknown. As they contain a closed internal fermion (vortex line)
loop we here assume a negative sign.

\paragraph{The identification with the Odderon}

Even though QCD cannot presently be used to calculate soft processes the typical
absence of abrupt changes in experimental distributions indicates that there
is no discontinuous transition between soft and hard reactions both formulated
on a partonic level in the frame work of the topological model. This provides
the hope that hard processes can be used as a guide and that soft processes
can be parametrized as an extrapolation of calculable hard processes. 

The topological considerations are based on the \( 1/N_{C} \) - expansion.
Not to loose some of the younger physicist, this approximations selects contributions
according to the magnitude of their color factors. To leading order \( 1/N_{C} \)
gluons can be represented by pairs of color lines and the color factors just
represent the number of coloring choices. For an amplitude of a given structure
with a given number of couplings the leading contribution can be drawn without
crossing. An example of a leading and a nonleading contribution is shown below:

{\par\centering \resizebox*{0.33\textwidth}{!}{\includegraphics{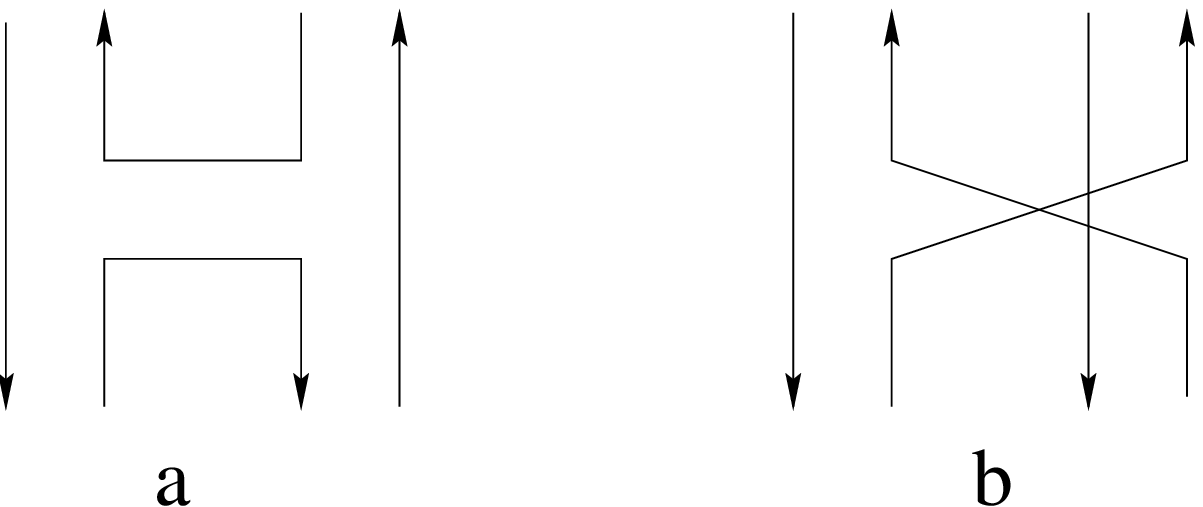}} \par}

\noindent The leading term (denoted ``a'') contains a new line on the top
whose color can be freely chosen. Some amplitudes require special contributions;
in this situation the leading terms can be drawn without crossing on topological
structures which are more complicated than the simple plane considered above.
An example is the cylinder assumed to be responsible for the Pomeron contribution.

\noindent The known example of the soft hard correspondence is the connection
between soft and hard Pomerons. To identify the hard partner of the soft Pomeron
we first observe that the simplest representation of a Pomeron in PQCD involves
the exchange of two gluons which can form color singlets with the required positive
charge parity. Following this concept it can be shown \cite{31} that a generalization
of such an exchange gives the dominant contribution at very high energies in
a well defined approximation. It is called ``hard'' or BFKL Pomeron and involves
a ladder of two exchanged Reggeized gluons linked by a number of gluons. In
the topological expansion the leading structure of a BFKL Pomeron corresponds
to a cylinder with the two basic gluons exchanged on opposite sides parallel
to the axis: 

{\par\centering \resizebox*{0.33\textwidth}{!}{\includegraphics{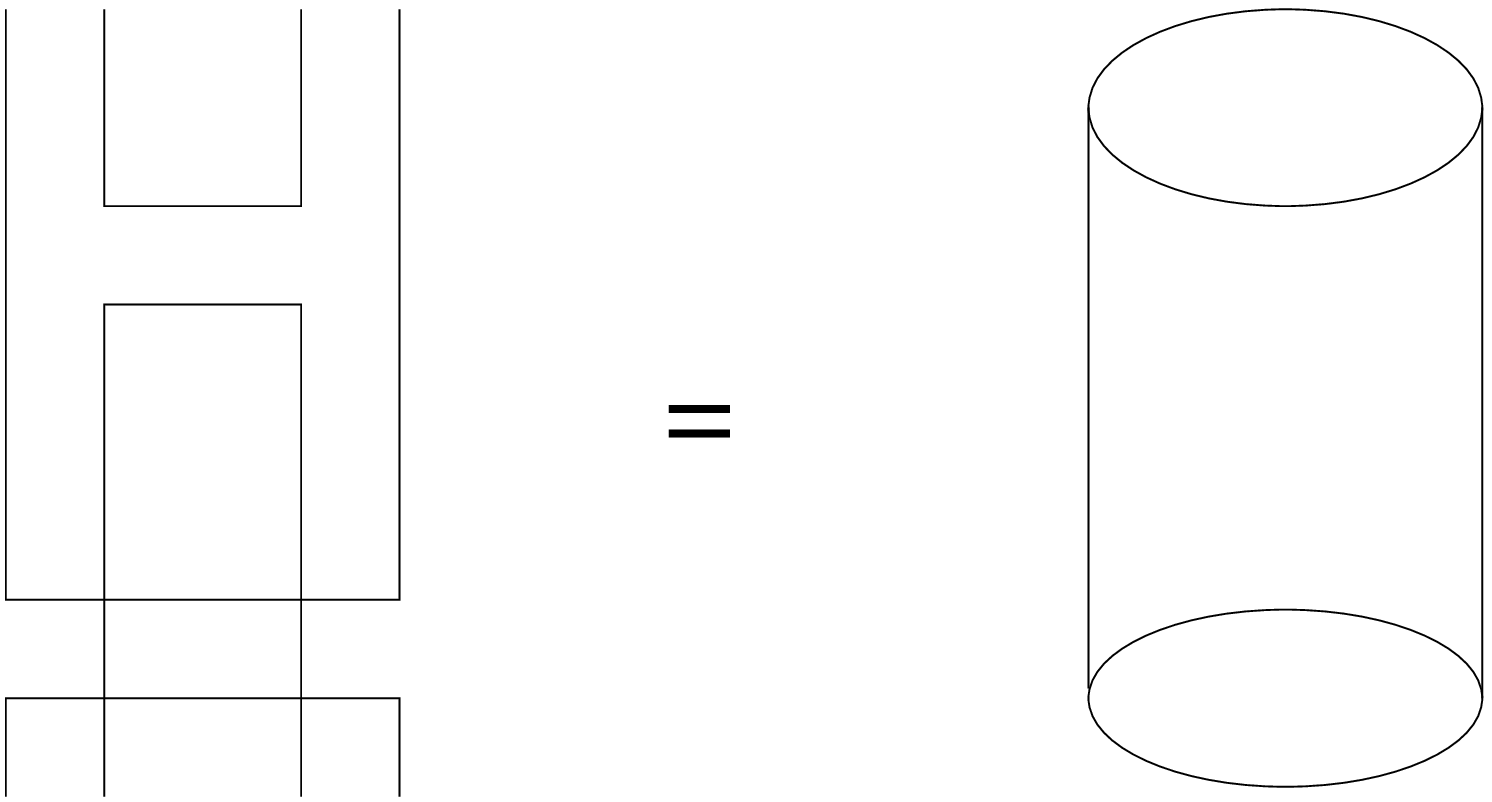}} \par}

\noindent Their matching inner color lines can be linked in front of the cylinder
without color line crossing. Analogously their matching outer lines can be connected
on the back of the cylinder. 

Going back to the soft regime the basic assumption in topological models is
that the \( 1/N \)-expansion stays valid and that the soft Pomeron therefore
maintains its cylindrical structure needed for the two string phenomenology
of hadronic final states. If cut, soft and hard Pomerons therefore lead to similar
two string final states. As difference it remains that the trajectory of the
observed soft Pomeron is just shifted downward roughly by a third of a unit
from hard Pomeron calculated in leading logarithmic approximation.

Can one find a similar connection for the membraned cylinder? The simplest representation
spanning such a topological structure involves three gluons, one on each sheet
exchanged parallel to the axis. Any gluon linking these exchanges has then to
pass through a vortex line in which the three sheets join. In the \( 1/N \)
expansion extended to baryons this means that the color lines have to cross
passing this line. The basic structure of the membraned cylinder exchange is
therefore the following:

{\par\centering \resizebox*{0.33\textwidth}{3cm}{\includegraphics{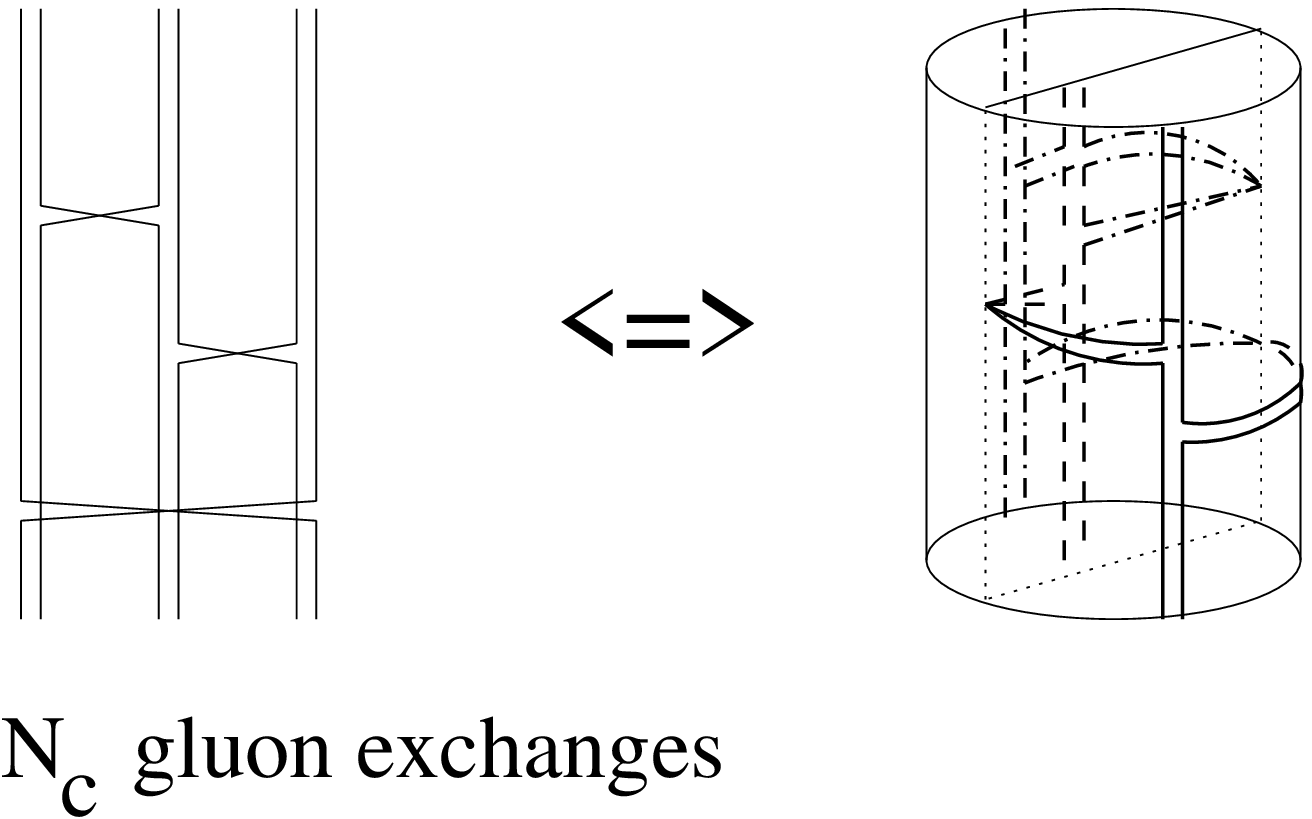}} \par}
\vspace{0.3cm}

Looking from the other side a color singlet of three gluons can have the quantum
numbers of a Pomeron or an Odderon \cite{32}. There is a simple topological
property of the Odderon. A single uncrossed gluon connection (of type ``a'')
would project the color structure of the pair to that of a single gluon and
the exchange would have to correspond to a Pomeron-like contribution. The Odderon
will therefore have to involve crossed links. Hence it has exactly the topology
of the membraned cylinder.

To visualize the baryonic color structure of the Odderon with its crossed exchanges
one can replace the exchanged gluons by quark antiquark pairs without changing
color lines. The so modified membraned cylinder just represents an exchanged
baryon antibaryon pair.

In the same QCD approximation as the ``hard'' Pomeron the properties of a ``hard''
or BKP Odderon \cite{33} were calculated and the predicted intercept is \( 0.96 \)
\cite{34}. Again a mismatch between this hard leading logarithmic Odderon intercept
and the experimentally observed soft value (preliminarily \( \approx 0.6\pm 0.2 \))
by about a third of a unit can be expected the Odderon trajectory.

\section{Experimental consequences \\
 of membraned cylinder exchanges.}

\textbf{Odderon in fits to total hadronic cross sections}

\noindent The assumed negative contribution from the asymmetric two string cut
makes the understanding of total cross sections difficult on a quantitative
level. It is possible that the membraned cylinder exchange has a small or almost
vanishing imaginary part. In this way there are no (presumably anyhow not serious)
constraints from total cross section fits. \textcolor{black}{The cancellation
allows a small or vanishing Odderon} \textcolor{black}{to contain clearly measurable
individual components. In this way data on baryon exchange can be used to determine
the }

\paragraph{Odderon in heavy ion scattering \newline}

\noindent In heavy ion scattering where the Pomerons are dense in transverse
space they can join and form a Pomeron or a membraned cylinder. The individual
strings are no longer independent but the general picture of particle production
in separate universal strings survives. The probability of an interaction of
strings and of membraned cylinder exchanges is growing proportional to the density:
\begin{equation}
\frac{{[}number\: of\: membraned\: cylinders{]}}{{[}Pomeron\: number{]}}\propto \frac{{[}Pomeron\: number{]}{[}Pomeron\: radius{]}^{2}}{{[}nucleus\: radius{]}^{2}}
\end{equation}

The transition from a Pomeron pair to the centrally cut membraned-cylinder involves
baryon antibaryon pair production. Between a proton and a Pomeron the cut membraned-cylinder
is a very efficient mechanism of baryon stopping. Both effects correspond to
experimental observations. As the trajectory is not well determined it is hard
to obtain really reliable quantitative statements which can be tested convincingly
with results in heavy ion scattering.

\paragraph{The backward peak in diffraction and possibly in electro production}

There is however a very specific qualitative prediction which can be tested.
Consider a diffractive system whose mass exceeds ISR energies. Usually the diffractively
produced particles will originate in two strings of a cut Pomeron and the baryon
charge will stay on the side of the initial proton. As usual there might be
some migration to the center with a slope in rapidity eventually corresponding
to the difference of the Odderon and the Pomeron trajectory. Topologically it
involves a horizontal cut through the following structure:

\vspace{0.3cm}
{\par\centering \resizebox*{!}{3cm}{\includegraphics{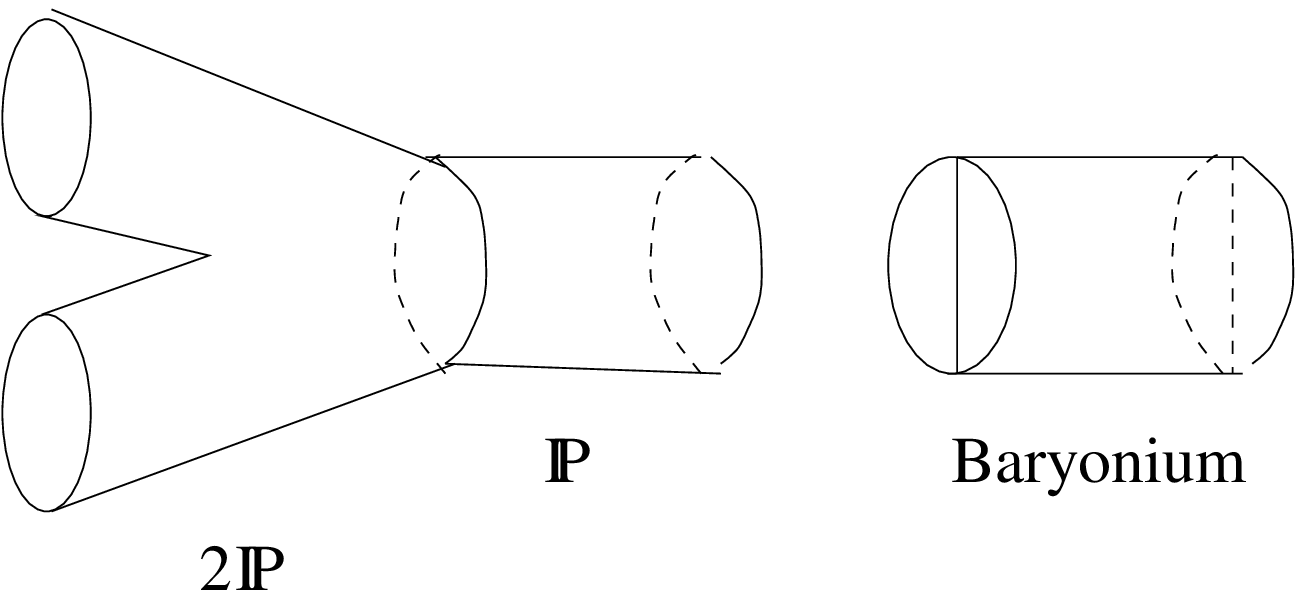}} \par}
\vspace{0.3cm}

The high Odderon trajectory argued for above requires a clear suppression from
the coupling constants to stay consistent with data. No such suppression is
expected at a two Pomeron vertex. In consequence at a certain distance it should
be more favorable for the membraned cylinder to span the total diffractive region
and to utilize the more favorable coupling to the two Pomerons. In this way
the initial baryon will sometimes end up exactly at the end of the string. It
should be visible if one plots the rapidity distribution in relation to the
inner end of the diffractive region, i.e. as function of 

\[
y_{\{Pomeron\}}=y_{\{CMS\}}-\ln \frac{m\sqrt{{s}}}{M(diffr.)}\]

To illustrate the expected small backward peak we show the result of a calculation
with the PHOJET Monte Carlo code \cite{35} of the incoming proton spectrum
for diffractive events with a mass of \( 300 \) GeV for \( pp \)-scattering
of \( 1.8 \) TeV with standard parameters below. To select diffractive events
a lower cutoff of \( x_{F}=0.95 \) was used.
\begin{figure}
{\par\centering \includegraphics{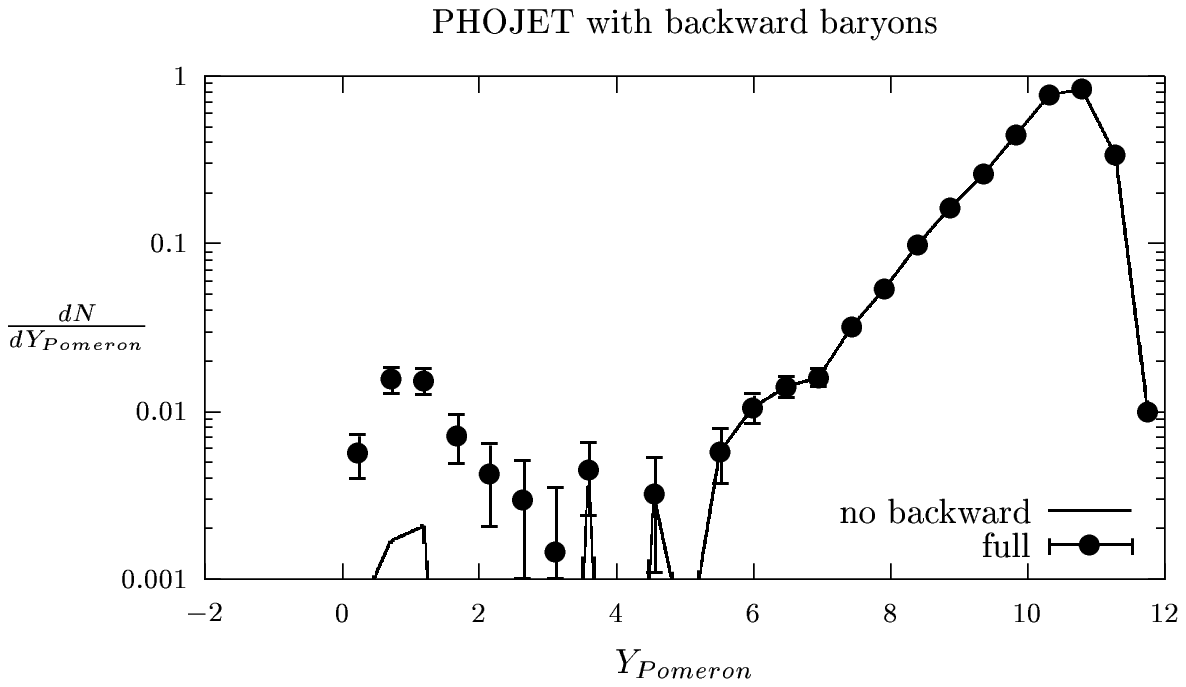} \par}
\end{figure}
PHOJET contains diquark exchanges and yields reasonable baryon spectra in the
forward region. To obtain the postulated backward peak we just mixed in a suitable
sample of inverted events (with disabled diquark exchanges).

\section{Conclusion}

Our aim with this talk and with the more detailed paper \cite{36} is to encourage
measurement of the initial baryon distribution in high mass diffractive systems.
\textcolor{black}{Similar measurements of a backward peak in electro production
might also be possible at HERA.} 

The prediction is important as it has manifest consequences for heavy ion processes,
where it offers a strong mechanism for central baryon production and for the
transport of initial baryons to the central and opposite region. It might also
clarify the role of the Odderon.

\end{document}